\documentstyle[aps,prl]{revtex}
\begin{document}
\bibliographystyle{unsrt}

\draft
\title{Manifestation of classical bifurcation in the spectrum
of the integrable quantum dimer
} 
\author{S. Aubry}
\address{Laboratoire L\'{e}on Brillouin, CEN Saclay, 91191 Gif-sur-Yvette,
France}
\author{S. Flach, K. Kladko and E. Olbrich}
\address{Max-Planck-Institute for Physics of Complex Systems, Bayreuther
Str. 40 H.16, D-01187 Dresden, Germany}
\date{\today}
\maketitle
\begin{abstract}
We analyze the classical and quantum properties of the integrable dimer
problem. The classical version exhibits exactly one bifurcation 
in phase space, which gives birth to permutational symmetry 
broken trajectories and a separatrix. The quantum analysis yields
all tunneling rates (splittings) in leading order of perturbation. 
In the semiclassical regime the eigenvalue spectrum
obtained by numerically exact diagonalization allows 
to conclude about the presence of a separatrix
and a bifurcation in the corresponding classical model. 
\end{abstract}


The problem of correspondence between classical and quantum-mechanical
properties of nonlinear systems is currently an object of wide
interest \cite{mcg90}. One interesting topic concerns Hamiltonian systems with
a given symmetry (e.g. some permutational symmetry), where classical
trajectories exist which are not invariant under the corresponding 
symmetry operation. This topic appears in analyzing selective bond
excitation in chemistry and in the quantization of discrete breathers
\cite{am95}.

We consider an integrable system with two degrees of freedom (TDF), 
whose classical version exhibits exactly one bifurcation (of periodic 
orbits) and separatrix manifold. This manifold cuts the phase space into three
parts - one with invariant trajectories, and two with noninvariant
trajectories, where the corresponding symmetry is the permutational
one. By varying a single parameter it is possible to 'switch' between
these phase space parts by crossing the separatrix. It appears natural
to expect in the quantum case a drastic change in the splittings of
energy levels (which should be zero in the classical limit for the
noninvariant phase space parts). However the splittings are nonzero
for any given value of the control parameter. The only way to avoid
contradiction between the classical and quantum cases is to assume that
the quantum level splittings tend to a step-like function (of e.g.
the level pair number) in the classical limit. The step should occur
at the position of the classical separatrix.
This problem can be also
coined {\sl dynamical tunneling } through a separatrix. There exist
studies of the influence of classical chaos on dynamical tunneling
\cite{btu93}, however
we are not aware of any systematic study in the {\sl absence} of chaos.

We are able to trace the splittings of the level pairs using {\sl quantum}
perturbation methods. We consider the quasiclassical regime and show
that the step indeed occurs. Therefore we are able to extract information
about the classical separatrix and bifurcation. Further we 
show, that the quantum density of states (the second integral of motion is 
fixed) exhibits a sharp maximum at the separatrix energy. By calculating the
corresponding classical quantity (with the help of Weyl's formula)
we find that this singularity appears due to the integration over a part
of the separatrix manifold which includes a hyperbolic isolated orbit.

Let us consider the integrable dimer model
with Hamiltonian \cite{els85}
\begin{equation}
H=\frac 12\left(P_1^2 + P_2^2 + X_1^2 + X_2^2\right) +
\frac 18\left( (P_1^2 +X_1^2)^2 + (P_2^2 +X_2^2)^2 \right)
+ \frac C2\left( X_1X_2 + P_1P_2 \right) \;\;. \label{1-1}
\end{equation}
Here $P_{1,2},X_{1,2}$ are canonically conjugated momenta and positions
of two degrees of freedom. System (\ref{1-1}) is integrable, because
the classical Poisson bracket of 
\begin{equation}
B = P_1^2 + P_2^2 + X_1^2 + X_2^2 \label{1-2}       
\end{equation}
with $H$ vanishes. Further (\ref{1-1}) is invariant under permutation of
indices.

With $\Psi=1/\sqrt{2}(X + iP)$ (\ref{1-1}) becomes
\begin{equation}
H = \Psi^*_1\Psi_1 + \Psi^*_2\Psi_2 + \frac 12 \left( (\Psi^*_1\Psi_1)^2
+ (\Psi^*_2\Psi_2)^2 \right) + C \left( \Psi^*_1\Psi_2 + \Psi^*_2\Psi_1
\right) \;\;. \label{1-3}       
\end{equation}
The equations of motion become $\dot{\Psi}_{1,2} = i\partial H / \partial
\Psi^*_{1,2}$.

Isolated periodic orbits (IPO)
satisfy the relation ${\rm grad}H \;\; ||\;\; {\rm grad}B$.
Let us parametrize the phase space of (\ref{1-3}) with $\Psi_{1,2}=A_{1,2}
{\rm e}^{i\phi_{1,2}}$, $A_{1,2} \geq 0$. 
It follows $A_{1,2}$ time independent and $\phi_1=\phi_2 + \Delta$ with
$\Delta=0,\pi$ and
$\dot{\phi}_{1,2}=\omega$ time independent. Solving the algebraic equations
for the amplitudes of the IPO's we obtain
\begin{eqnarray}
{\rm I:}\; A_{1,2}^2 = \frac 12 B \;,\; \Delta=0\;,\;
\omega=1+C+\frac 12 B \;\;,\label{1-4a} \\
{\rm II:}\; A_{1,2}^2 = \frac 12 B \;,\; \Delta=\pi\;,\;
\omega=1-C+\frac 12 B \;\;,\label{1-4b} \\
{\rm III:}\; A_1^2=\frac 12 B\left(1\pm \sqrt{1-4C^2/B^2}\right)\;,\;
\Delta=0\;,\;\omega = 1+B\;\;.\label{1-4c}
\end{eqnarray}
IPO III corresponds to two elliptic solutions which break the permutational
symmetry. IPO III exist for $B \geq B_b$ with $B_b=2C$ and occur through
a bifurcation from IPO I \cite{els85}. 
The corresponding separatrix manifold is uniquely
defined by the energy of IPO I at a given value of $B \geq B_b$. This manifold
separates three regions in phase space - two with symmetry broken solutions,
each one containing one of the IPO's III, and one with symmetry conserving
solutions containing the elliptic IPO II. The separatrix manifold 
itself contains the
hyperbolic IPO I. For $B \leq B_b$ only two IPO's exist - IPO I and II, with
both of them being of elliptic character. Remarkably there exist no other
IPO's, and the mentioned bifurcation and separatrix manifold are the only
ones present in the classical phase space of (\ref{1-1}) \cite{els85}.

To conclude the analysis of the classical part, we calculate the energy
properties of the different phase space parts separated by the separatrix
manifold. First it is straightforward to show that the IPO's 
(\ref{1-4a})-(\ref{1-4c}) correspond to maxima, minima or saddle points
of the energy in the allowed energy interval for a given value of $B$, with
no other extrema or saddle points present \cite{els85}. It follows 
\begin{eqnarray}
E_1=H({\rm IPO\; I}) = B + \frac 14 B^2 + CB \;\;,\label{1-5a} \\
E_2=H({\rm IPO\; II}) = B + \frac 14 B^2 - CB \;\;,\label{1-5b} \\
E_3=H({\rm IPO\; III}) = B + \frac 12 B^2 + C^2 \;\;. \label{1-5c}
\end{eqnarray}
For $B < B_b$ we have $E_1 > E_2$ (IPO I - maximum, IPO II - minimum). For
$B \geq B_b$ it follows $E_3 > E_1 > E_2$ (IPO III - maxima, IPO I - saddle,
IPO II - minimum). If $B < B_b$, then all trajectories are symmetry conserving.
If $B \geq B_b$, then trajectories with energies $E_1 < E \leq E_3 $ are 
symmetry breaking, and trajectories with $E_2 \leq E \leq E_1$ are symmetry
conserving.

The quantum eigenvalue problem can be properly analyzed in second
quantization, which amounts to replacing the complex functions $\Psi, \Psi^*$
in (\ref{1-3}) with the boson annihilation and creation operators 
$a,a^+$ with standard commutation relations
(to enforce invariance
under exchange $\Psi \Leftrightarrow \Psi^*$ the substitution has to be
done after rewriting
$\Psi\Psi^*=1/2(\Psi \Psi^* + \Psi^* \Psi)$):
\begin{equation}
H = \frac{5}{4} +  \frac{3}{2}\left( a_1^+a_1 + a_2^+a_2 \right)  
+ \frac 12 \left( (a_1^+a_1)^2 + (a_2^+a_2)^2 \right)
+ C \left( a_1^+a_2 + a_2^+a_1 \right) \;\;. \label{1-6}       
\end{equation}
Note that $\hbar = 1$ here, so the eigenvalues $b$ of $B=a_1^+a_1 + a_2^+a_2$ 
are integer numbers. Since $B$ commutes with $H$ we can diagonalize the 
Hamiltonian in the basis of eigenfunctions of $B$. Each value of $b$ spans
a subspace of dimension $(b+1)$ in the space of eigenfunctions. 
These eigenfunctions
are products of the number states $|n>$ of each degree of freedom and
can be characterized by a symbol $|n,m>$ where we have $n$ bosons on
site 1 and $m$ bosons on site 2. For a given value $b$ it follows $m=b-n$.
So we can actually label each state by just one number $n$: $|n,(b-n)>
\equiv |n)$.
Consequently the eigenvalue problem at fixed $b$ amounts to diagonalizing
the matrix $H_{nm}$ with
\begin{equation}
H_{nm}= \left\{
\begin{array}{lr}
\frac{5}{4} + \frac{3}{2}b + \frac{1}{2}\left( n^2 + (b-n)^2
\right) & n=m \\
C\sqrt{n(b+1-n)} & n=m+1 \\
C\sqrt{(n+1)(b-n)} & n=m-1 \\
0                  & {\rm else} 
\end{array}
\right 
. \label{1-8}
\end{equation}
and $n,m=0,1,2,...,b$.
Notice that the matrix $H_{nm}$ is a symmetric band matrix. The additional
symmetry $H_{nm}=H_{(b-n),(b-m)}$ is a consequence of the permutational
symmetry of $H$.

For $C=0$ the matrix $H_{nm}$ is diagonal, with the property that
each eigenvalue is doubly degenerated (with exception of the state
$|b/2)$ for even values of $b$). 
The classical phase space contains only symmetry broken
trajectories, with the exception of IPO II and the separatrix with IPO I
(in fact in this limit the separatrix manifold is nothing but a resonant
torus containing both IPO's I and II). So with the exception of the separatrix
manifold, all tori break permutational symmetry and come in two groups
separated by the separatrix. Then quantizing each group will lead to
pairs of degenerated eigenvalues - one from each group. There is a clear
correspondence to the spectrum of the diagonal ($C=0$) matrix $H_{nm}$.
The eigenvalues $H_{00}=H_{bb}$ correspond to the quantized IPO's III.
With increasing $n$ the eigenvalues $H_{nn}=H_{(b-n),(b-n)}$ correspond
to quantized tori further away from the IPO III. Finally the states
with $n=b/2$ for even $b$ or $n=(b-1)/2$ for odd $b$ are tori most close
to the separatrix.
Switching the side diagonals on by
increasing $C$ will lead to a splitting of all pairs of eigenvalues.
In the case of small values of $b$ these splittings have no correspondence
to the classical system properties. However in the limit of large $b$
we enter the semiclassical regime, and due to the integrability of the 
system eigenfunctions should correspond to tori in the classical phase
space which satisfy the Einstein-Brillouin-Keller quantization rules
\cite{mcg90}. 

Increasing $C$ from zero will lead to a splitting $\Delta E_n$ of the eigenvalue
doublets of $C=0$. In other words we find pairs of eigenvalues, which
are related to each other through the symmetry of their eigenvectors and
(for small enough $C$) through the small value of the splitting. 
Let us calculate the splittings in leading perturbation order. 
This is done by applying standard perturbation
theory to each of the states $|n)$ and $|(b-n))$ and calculating the
perturbed eigenvectors until the matrix element of the two perturbed
eigenvectors with $H$ does not vanish. 
Due to the band structure of our matrix the final result has the following
form \cite{bga94}:
\begin{equation}
\Delta E_n = 2 \prod _{i=n}^{b-n-1}H_{i,(i+1)} \prod_{i=n+1}^{b-n-1}
\left( H_{nn} - H_{ii} \right)^{-1}\;\;.\label{1-9}           
\end{equation}
For even $b$ with $\tilde{n}=n-b/2$ and (\ref{1-8})
it follows
\begin{equation}
\Delta E_n = 2 C^{2|\tilde{n}|} \frac{(\frac{b}{2} + |\tilde{n}|)!}
{(2|\tilde{n}|-1)!^2 
(\frac{b}{2} - |\tilde{n}|)!} \;\;. \label{1-10a}     
\end{equation}
For odd $b$ with $\tilde{n}=n-b/2 + 1/2{\rm sgn}(n-b/2) $ and (\ref{1-8})
we find
\begin{equation}
\Delta E_n = 2 C^{2|\tilde{n}|-1}\frac{(\frac{b-1}{2}+|\tilde{n}|)!}
{(2|\tilde{n}|-2)!^2 (\frac{b+1}{2}-|\tilde{n}|)!}\;\;.\label{1-10b}
\end{equation}
The integer $\tilde{n}$ counts the pairs of equal diagonal elements of 
(\ref{1-8}) from the center of $H_{nm}$ towards the corners ($b$ even:
$|\tilde{n}|=0,1,2,...,b/2$ and $b$ odd: $|\tilde{n}|=1,2,...,(b+1)/2$).
Note that for the corner states the obtained expression for the splitting
is identical with the results in
\cite{bes90}. 
Let us define
$|\tilde{n}|=\alpha b/2$ with $0 < \alpha < 1$.
For fixed $\alpha$ application of Stierling's
formula to (\ref{1-10a}),(\ref{1-10b}) yields
\begin{equation}
\Delta E_n \approx \frac{b}{\pi {\rm e}} \left( \frac{1+\alpha}{1-\alpha}
\right)^{1/2}  \gamma 
^{\alpha b}\;,\; \gamma =
\frac{{\rm e}C \sqrt{1-\alpha^2}}{2\alpha(\alpha b -1)}
\left( \frac{1+\alpha}{1-\alpha}
\right)^{1/(2\alpha)}\;. \label{1-11}       
\end{equation}
For large $\alpha b$ the expression (\ref{1-11}) should be close to
zero if $\gamma < 1$ and its inverse should be close to zero if
$\gamma > 1$. So the perturbation result predicts a step-like change
in the splitting values for $\gamma=1$ in the limit of large $\alpha b$.
The considered asymptotic limit corresponds to the classical limit of
(\ref{1-6}). Thus we expect that the splittings of the eigenvalue pairs
which correspond to symmetry broken classical tori should vanish in this limit. 
Consequently the condition $\gamma = 1$ predicts the position of the
classical separatrix with respect to the variable $\alpha$.

Now we calculate the eigenvalue spectrum of (\ref{1-6}) numerically\footnote{ 
This was done
using standard Fortran routines with double precision. When splittings
had to be calculated with values below $10^{-16}$ we used Mathematica
routines, where the precision can be of any value \cite{sw91}.}
(for $b=20$ this was done in \cite{ljb93}). 
In Fig.1 we show the eigenvalues (grouped with respect to
their eigenfunctions being symmetric or antisymmetric with respect to
permutation)  as a function of $\tilde{n}$
for $b=600$ and $C=50$. The classical model has symmetry broken trajectories,
and a separatrix with energy $E_{sep}=E_1=120600$. For the quantum problem
we find an inflection point in the eigenvalue spectrum of each subgroup
at precisely this energy ($\tilde{n} \approx 150$). Since $\tilde{n}(E)$
is the integrated density of states, its derivative with respect to $E$
gives the density of states $\rho (E)$, which hereby exhibits a peak at the 
separatrix energy of the classical system (inset in Fig.1). Using Weyl's formula
we can calculate its classical counterpart \cite{mcg90} 
\begin{equation}
\rho_{cl}(E,b)= \int {\rm d}^2P {\rm d}^2X \delta \left (E-H(P,X) \right) \delta
\left (b-B(P,X) \right) \;\;. \label{1-12}         
\end{equation}
This integral can be rewritten as $\rho_{cl}(E,b)=\oint1/(|\nabla H|
|\nabla B| {\rm sin}\Theta) {\rm d}S $, where the integration is done over the
surface of constant $H$ and $B$ and $\Theta$ is the angle between the
two gradients. The denominator vanishes on IPOs. Expanding the denominator
in a Taylor series in the neighbourhood of an IPO it follows, that for
elliptic IPOs no singularity develops (because the torus surface vanishes)
whereas for hyperbolic IPOs (i.e. on the separatrix) a logarithmic
singularity appears. 

By parametrizing the classical phase space using $A_{1,2}$ and $\phi_{1,2}$
the expression (\ref{1-12}) can be reduced to a single integral:
\begin{equation}
\rho_{cl}(E,b) = \frac{1}{\pi}\int \frac{{\rm d}y}{\sqrt{
C^2b^2-4C^2y^2-\left( E-b-b^2/4 - y^2\right) ^2}}\;\;. \label{1-13}   
\end{equation} 
The integration has to be done over all values of $y$ where the expression
under the root is nonnegative.
This integral shows up with a singularity at the classical separatrix
energy.
The numerical
integration is compared in the inset in Fig.1 
with the quantum density of states.
We find excellent agreement. In  the inset in Fig.2 the splittings are shown  
with respect to $\tilde{n}$. The splittings become anomalously small
in the region of classical symmetry broken solutions, which is bounded
again by the separatrix energy. In Fig.2 we compare the numerically
obtained splittings with the perturbation theory result ($b=150$,
$C=10$). Even though
the true splittings become as small as $10^{-100}$ compared to the 
averaged spacings, the perturbation theory reproduces at the best the
order of magnitude, but fails by e.g. 50\% in the absolute value.
Consequently we note that higher order terms in the perturbation theory
are important even when the true splittings are anomalously small.

Still there is useful information in the perturbation result as shown in
(\ref{1-11}). In Fig.3 we show the classical separatrix energy
$E_1$ for different values of $C$ ($b=600$) 
and compare it to the peak energy in the quantum
density of states {\sl and} to the condition $\gamma =1$ (which gives us
a certain $\alpha$, which in turn yields a given $\tilde{n}$ and through
the numerically obtained quantum eigenvalue spectrum a corresponding
energy). First we note the remarkable agreement between the classical
curve and the exact quantum counterpart. But even the perturbation
theory gives values which deviate by only 6\% from the exact result.
So while the perturbation theory fails in reproducing the absolute values
of the splittings, it still contains the information about a classical
separatrix with good precision. 

Finally we can easily trace the classical bifurcation by considering
the dependence of the largest eigenvalue of the quantum spectrum as 
a function of $C$: $E_{max}=f(C)$. According to the classical system
this function is given by (\ref{1-5a}) for $C > b/2$ and by (\ref{1-5c})
for $C< b/2$. Differentiating this function twice with respect to
$C$ should thus yield a step function with the step located at $C=b/2$.
In the inset of Fig.3 ${\rm d}^2f/{\rm d}C^2$ is shown for $b=600$.
The step at $C=300$ is nicely observed.
\\ \\ \\
Acknowledgements
\\
\\
We thank L. Bernstein, C. Eilbeck, H. Kantz, B. Mehlig, K. M\"uller 
and A. Scott
for valuable discussions.

FIGURE CAPTIONS
\\
\\
Fig.1: 
\\
Eigenvalues of the symmetric eigenstates (solid line) and
antisymmetric eigenstates (dashed line) versus quantum number
$\tilde{n}$ for $b=600$ and $C=50$.
\\
Inset; Density of states for the eigenvalue spectrum from above
(solid line) versus energy. The dashed line is the classical
prediction using Weyl's formula.
\\
\\
Fig.2:
\\
Eigenvalue splittings versus quantum number $\tilde{n}$ for
$b=150$ and $C=10$ (calculated with precision 512). Solid line -
exact diagonalization, dashed line - perturbation theory result.
Note that even for $\tilde{n} \approx 80$ the ratio of both values
is of the order of 0.5.
\\
Inset: Eigenvalue splittings versus $\tilde{n}$ for $b=600$
and $C=50$ (compare Fig.1) from exact diagonalization. Splittings
are of the order of average spacing for $\tilde{n} < 150$ and
collapse to zero for $\tilde{n}>150$.
\\
\\
Fig.3:
\\
Separatrix energy versus $C$ for $b=600$ for the classical system
(solid line). The thick long-dashed line is the position of the
maximum in the quantum density of states. the thin dashed-dotted line
is the perturbation theory prediction ($\gamma =1$).
\\
Inset: Second derivative of the $C$-dependence of the maximum
eigenvalue of the quantum spectrum for $b=600$ versus $C$. The
classical prediciton is a step function with values 2,0 and step position
$C=300$.

\end{document}